\documentclass[manuscript]{aastex}

\usepackage{graphicx}

\newcommand{\etal}{{\it et al}.\ }

\slugcomment{Submitted to ICARUS}

\shorttitle{Collisions, Radiation and the Trojans}

\shortauthors{Melita et al.}

\begin{document}

\title{Collisions, Cosmic Radiation and the Colors of the Trojan
Asteroids}

\author{M.D. Melita\altaffilmark{1}}
\affil{Instituto de Astronom\'ia y F\'isica del Espacio. Buenos Aires.
Argentina. }
\email{melita@iafe.uba.ar}

\and

\author{G. Strazzulla\altaffilmark{2}}
\affil{INAF-Osservatorio Astrofisico di Catania, Italy}
\email{gianni@oact.inaf.it}

\and

\author{A. Bar-Nun\altaffilmark{3}}
\affil{Department of Geophysics and Planetary  
Sciences.Tel Aviv University,Tel Aviv,Israel. }
\email{akivab@post.tau.ac.il}

\begin{abstract}

The Trojan asteroids orbit about the Lagrangian points of Jupiter and the
residence times about their present location are very long for most of them.
If these bodies originated in the outer Solar System, they should be mainly
composed of water ice, but, in contrast with comets, all the volatiles close
to the surface would have been lost long ago. Irrespective of the rotation
period, and hence the surface temperature and ice sublimation rate, a dust
layer exists always on the surface. We show that the timescale for
resurfacing the entire surface of the Trojan asteroids is similar to that of
the flattening of the red spectrum of the new dust by solar-proton
irradiation. This, if the cut-off radius of the size distribution of the
impacting objects is between 1mm and 1m and its slope is -3, for the entire
size-range. Therefore, the surfaces of most Trojan asteroids should be
composed mainly of unirradiated dust.
 
\end{abstract}


\section{Introduction}

All the Trojan asteroids orbit about the Sun at roughly the same
heliocentric distance as Jupiter, residing close to the Lagrangian stable
points, leading or trailing at $\sim 60^o$ from the planet. They are
particularly interesting because of being relatively isolated, lying just
beyond the {\it snow line}, the distance at which water can exist in the
form of ice. As such, it may be expected that the bulk of the Trojans were
mainly composed of ices.  However, at the moment, no water or other more
volatile substances have been detected on the surface of a Trojan (Yang \&
Jewitt 2007, Emery \& Brown 2003; 2004). In the case of Ennomos, an
unusually hight-albedo object, the surface-content of water ice has been
quantified to be below $10$\% in mass (Yang \& Jewitt 2007). However the
lack of detection does not imply that water ice is not there. Brunetto and
Roush (2008) have in fact shown that a relatively thin (tens of microns)
crust of specific refractory materials (they considered irradiated methane
ice) can mask the presence of water ice bands in the spectrum.

\begin{figure}
\epsscale{.80}
\plotone{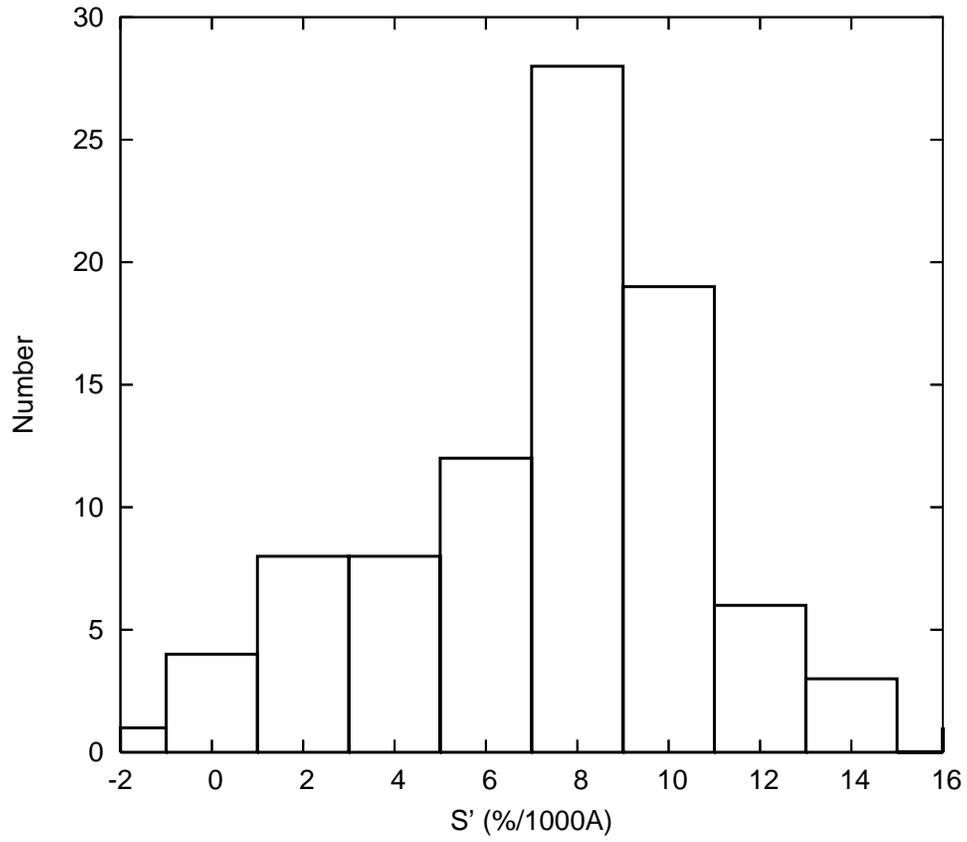} 
\caption{Histogram of the slopes of visual reflection spectra
normalized at $6000 \AA$ for the Trojan asteroids. The sources of this
data can be found in Melita et al. 2008.}
\label{fig:1}
\end{figure}

The Trojans have been characterized as similar to cometary nuclei or
extinct comets (Jewitt and Luu 1990). Most of the objects fall into
the D and P asteroid taxonomic classes and a few objects show spectra
with neutral or even negative slopes in the visual range (Bendoya et
al. 2004), with the majority having moderately-red surfaces (see
figure~\ref{fig:1}). They have been observed to posses low albedos
(Fernandez et al. 2003) of about 4$\%$ and infrared observations
indicate the presence of silicates (Cruikshank et al. 2001, Emery et
al. 2006) and depletion of water ice and volatiles (Emery \& Brown
2003; 2004).

Numerical models of the collisional evolution of the Trojan asteroids
(Marzari et al. 1997) are successful in explaining the present Trojan
population (Jewitt et al. 2000). The steep slope of the Trojan size
distribution at diameters larger than about $50-100$km is
essentially unaltered during the collisional evolution and must
reflect the formation process of these bodies. At smaller sizes,
collisions have produced a power law size distribution, with a slope
characteristic of collisionally relaxed populations. The formation of
prominent dynamical families in the Trojan swarms is a natural outcome
of the collisional process, and indeed a significant number of
families are known (see for example Beauge \& Roig 2001).

\begin{figure}
\epsscale{.80}
\plotone{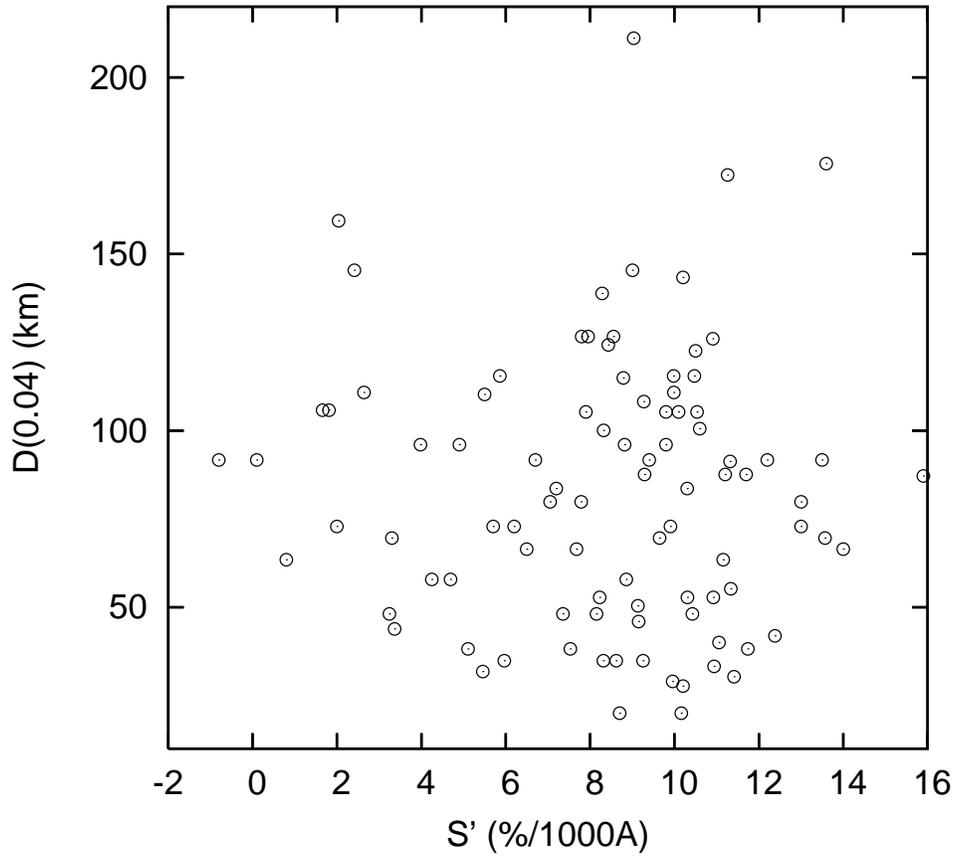}
\caption{Slopes of visual reflection spectra normalized
at $6000 \AA$ vs. D(0.04), the physical diameter, assuming an albedo
of 0.04, for the Trojan asteroids.
Data taken form Melita et al. 2008.}
\label{fig:2}
\end{figure}

It has also been found that the absolute-magnitude distribution of the
Trojans in stable orbits is bimodal, while that of the unstable orbits
is unimodal, with a slope similar to that of the small stable Trojans
(Melita \etal 2008a). This supports the hypothesis that the unstable
objects are mainly byproducts of physical collisions, the transitional
size has been found to be $H \approx 9.5$. But the slopes of the
reflectance spectra show no relation with the size of the objects (see
figure~\ref{fig:2}) or with the stability of the orbits (Melita \etal 2008a).

We intend to understand why most of the Trojans are moderately red,
regardless if they are large bodies or fragments of collisions. These
asteroids are subject to impacts which can extract fresh material from the
interior and scatter it on the surface. The fresh material would be 
composed of dust and ice. After the Deep-Impact event, we note the increase
of emission of both organic compounds and water ice (Lisse et al. 2006). But, if water ice were
deposited on the surface after an impact, it would sublimate very rapidly.
Therefore, after a physical impact, a layer of unirradiated ice --which has
a typically red spectral slope-- is deposited in the surface. It has been
suggested that terrestrial bitumens are good spectral analogues for such a
dust layer (e.g. Moroz et al 2004 and references therein). The less
energetic solar protons modifies the spectroscopic properties of this dust,
reducing its albedo and flattening the spectral slope (Moroz et al. 2004).
But if the collisional process can resurface the asteroids on a timescale
which is similar, or shorter, than the timescale that flattens the spectrum,
it is expected that most of the Trojans will be observed to be red. In the
following section we present a model to estimate relevant timescales and we
discuss its results in the last section.

\section{Surface alteration timescales}

\label{sec:sub}

\subsection{Timescale of flattening of the spectral slope by irradiation}

There are a number of energetic ion populations bombarding the
surfaces of small bodies in the outer Solar System: solar wind ions,
ions from solar flares, galactic cosmic rays, and the so called
"anomalous" cosmic ray component. The last component is however
relevant only to objects at the Pluto orbit and beyond.

Solar wind is an expanding flux of fully ionized plasma that reaches,
at distances greater than a few solar radii, an expansion speed of
about 400 km s$^{-1}$ corresponding to an energy of $\approx $1
keV/amu (slow solar wind, see e.g. Gosling 2007). At 1 AU its flux is
$\approx$10$^8$ protons cm$^{-2}$$s^{-1}$ and it decreases with the
square of the solar distance. The effects of the solar wind ions
irradiating "red" organic bitumens (namely asphaltite and kerite) have
been simulated in the laboratory (Moroz et al 2004). The hypothesis is
that terrestrial bitumens are good spectral analogues of the red
organic surfaces of small objects in the Solar System (we consider Trojans in
the present paper). The main result is a progressive flattening of the
red spectrum of bitumens with ion fluence. Moreover Moroz et al (2004)
suggest that the mechanism that produces the flattening is connected
to the energy deposited by elastic collision of the incoming ions with
the atoms in the target. This allows to extend the laboratory results
obtained with 10's keV H$^+$, N$^+$, and Ar$^{++}$ to all solar wind
ions. Because in the laboratory the flattening of the asphaltite
spectrum in the 0.3-0.8 micrometers spectral region has been obtained
after about 0.4 $\times$ 10$^{18}$ C-displacements cm$^{-2}$ (see
Moroz et al, top panel in Fig 14), we need to calculate the time
necessary to solar wind ions (at 5 AU) to cause the accumulation of
such an amount of damage on the bombarded surface. As said, solar wind
ions have energy of about 1 keV/amu. Most of them are protons (96 \%)
and He (4 \%) with traces of heavier elements. Using the TRIM code
(Transport of Ions in Matter; e.g., Ziegler et al. 1996) we calculate
that the number of displaced atoms in asphaltite is about 1.5 for 1
keV protons and 20 for 4 keV alpha particles. Considering the
respective abundances, we obtain an average of 2.2 C-displacements per
solar wind ion (the contribution of heavier ions can be neglected). At
a distance of 5 AU from the Sun we then expect that an exposed surface
will suffer about 10$^8$$\times$ 2.2/25= 9$\times$ 10$^6$
C-displacements cm$^{-2}$sec$^{-1}$. Thus the time necessary to
reproduce the effects observed in the laboratory is: 0.4$\times$
10$^{18}$/9$\times$10$^6$= 5.5$\times$ 10$^{10}$ sec i.e. about
2$\times$10$^3$ yrs.

It is important to note that solar wind
particle irradiation affects only a thin surface layer ($\leq$ 2000
$\AA$). This is however sufficient to affect the optical properties of
the surface (as simulated in the laboratory), but may be easily
destroyed by impact resurfacing. As discussed here, such 
resurfacing of irradiated material probably occurs constantly.

In order to produce a radiation-induced damage of deeper layers, more
energetic ions should be considered. Solar cosmic rays are produced
during solar flares. An average flux of solar protons having an energy
of 100 keV and a penetration depth of about 1 $\mu$m has been
estimated to be $\approx$ 2$\times$10$^{11}$ protons cm$^{-2}$
yrs$^{-1}$ (Lanzerotti et al., 1978). It is inversely proportional
both to the square of the energy and to the square of the solar
distance. At 5 AU the time necessary to reach the maximum
C-displacements of our experiments is about 2$\times$ 10$^7$ yrs.

Galactic cosmic rays are less abundant but significantly more
penetrating. In a recent review Hudson et al. (2008, and references
therein) reported the estimate of the radiation doses for objects at
different distances from the Sun (see their Table 1). It has been
found that objects orbiting at 5-35 AU accumulate in about 10$^9$ yrs
an irradiation damage sufficient to be fully chemically altered down
to depth of the order of 1 meter. Although galactic energetic
particles transfer most of their energy through electronic loss
(ionizations and excitations), however at the end of their path they
slow down and are thought to produce, in the so called Bragg peak, the
same effects meassured in the laboratory for low energy particles. 

In summary we have shown that the uppermost layers of an
object orbiting at the Jupiter distance can be weathered in about
10$^3$ yrs. If its surface is an organic material spectrally similar
to bitumens, its red sloped spectrum is flattened on that time scale.
Deeper layers are weathered in longer times: at a depths of the order
of 1 meter the target is fully altered in about 10$^9$ yrs.

\subsection{Timescale of sublimation of amorphous water-ice}

\label{sec:2.3}

The timescale for sublimation-mantle growth, $\tau_{M}$, is (Jewitt 2002):
\[\tau_{M} = \frac{\rho\ L}{f_M\ dm/dt},\] where $L$ is the depth of
the sublimated material, $\rho$ is the mass density of the asteroid,
$f_M$ is the fraction of the solid mass that cannot be ejected by gas
drag and $dm/dt$ is the drag mass-loss rate per unit area.

%The thermal-diurnal skin depth is $L_D = (\kappa\ P / \pi)^{1/2}$, where $P$
%is the spin period of the asteroid and $\kappa \approx 10^{-7}\ m^2s^{-1}$,
%is the thermal diffusivity of amorphous water ice (Bar-Nun et al. 2007). For
%a fast rotator, the effective temperature at the surface of a Trojan is
%$\sim 120^o K$ -assuming an albedo of $0.04$. At this temperature the mass
%loss rate of perfectly absorbing water-ice in equilibrium sublimation is
%$dm/dt \approx 3\ 10^{-10} kg\ m^{-2}\ s^{-1}$. 
%This figure can easily be
%derived from the water vapor pressure vs. Temperature value, taken from 
%CRC-Handbook of Chemistry 
%and Physics. 

For a fast rotator, the effective temperature at the surface of a Trojan is
$\sim 120^oK$ -assuming an albedo of $0.04$.  At this temperature, the mass
loss rate of perfectly absorbing water-ice in equilibrium sublimation is
approximately $dm/dt \approx 3\ 10^{-11} kg\ m^{-2}\ s^{-1}$. To find a
reliable estimation for $dm/dt$ at $T=120^oK$ is quite a complicated problem
and we have reached this value after much deliberation,
following the comments of the reviewer. The water-vapor pressure in $torr$, $P$,
vs. temperature, $T$, was calculated, using the polynomial from the
International Critical Tables between $140^oK$ and $180^oK$: 
$$\log \left( P \right) =
\frac{-2445.5646}{T} + 8.23121 \times \log(T) - 
0.167706 \times T + 1.20514 \times 10^{-5} T^{2} - 6.757169 , $$
and checked against the $P$ vs. $T$ in the CRC-Handbook of Chemistry and
Physics and found to agree with Sack and Baragiola (1993). The pressure at
$120^oK$ is $1.8 \times 10^{-10}\ Pa$. Since the flux of
sublimating water molecules, $F$, is, 
$$F=\frac{P(T)}{(2 \pi K m T)^{1/2}},$$ 
where $K$ is Boltzmann's constant and $m$ the mass of the molecule. We
recalculated the square root term in the denominator and found it to be: $(2 
\pi K m T)^{1/2} = 1.77 \times 10^{-23} kg\ m^{-1}\ s^{-1}$ per molecule.
Hence the flux calculated from the pressure as obtained by the polynomial
is: 
$$ F(T=120^oK)= \frac{P(T=120^o K)}{(2 \pi K m T)^{1/2}} = 1.1 \times
10^{13} molec.\ m^{-2}\ s^{-1} = 1.1 \times 10^{9}\ molec.\ cm^{-2}\
s^{-1}.$$ This flux was compared with the flux meassured by Ntesco and
Bar-Nun (2005) in their Figure 1 and we realized that the measured value at
$120^oK$ is the background of water vapor in our vacuum chamber.
Extrapolating the flux vs T plot in their Figure 1 to $120^oK$ leads to a
flux of approximately $10^{11}\ molec.\ cm^{-2}\ s^{-1}$, which is $100$
times larger than the one calculated by the P vs T through the polynomial,
due to the very large surface area of the amorphous ice, much larger than
the area calculated by the geometric dimension of the substrate on which the
ice was deposited. As measured by Bar-Nun et al. (1987), by Ar adsorption at
$120^oK$, a surface area of $38 m^2\ g^{-1}$ was found, in good agreement
with Mayer and Pletzer (1986) who measured $40 m^2\ g^{-1}$ at $114^oK$.
Sack and Baragiola' s (1993) Figure 2 data can be extrapolated to $8 \times
10^{10} molec.\ cm^{-2}\ s^{-1}$ at $120^oK$ and shows an $8$ fold increase
in the vapor pressure in amorphous non annealed ice, which they attribute to
the larger surface area. These authors (Figure 3)show a $7$ fold increase in
the flux at $135^oK$ with an ice thickness between $0.12$ and $5.1$ $\mu m$
and a $7-8$ fold decrease at $135^oK$ during $200 min$ at that temperature.
Both observations point to a considerable effect in surface roughness. To
be on the conservative side we shall adopt a flux of $5\ 10^{10}\ molec.\
cm^{-2}\ s^{-1}$ or $1.5\ 10^{-11} kg\ m^{-2}\ s^{-1}$. It should be
remembered that slight deviations from an albedo of $0.4$ or the fast
rotator approximation could result in large changes in the rate of
sublimation. The question is then which flux of water molecules from an ice
surface at $120^o K$ should be adopted?. The one obtained from Notesco and
Bar-Nun's (2005), $10^{11}\ molec.\ cm^{-2}\ s^{-1}$ or Sack and Baragiola's
(1993) extrapolated $8 \times 10^{10}\ molec.\ cm^{-2}\ s^{-1}$?. Since this
mass loss rate is very low, even the smallest sub-micron sized grains on the
surface of the Trojan asteroids are retained, hence, for the fast rotator
case, we take $f_M = 1$.

The known rotation periods of Trojans range approximately from $4 hr$ to $40
hr$ (Hartmann et al. 1988, Binzel and Sauter 1992, Melita et al.
2008b)\footnote{Rotation periods of Jupiter Trojan asteroids, determined
before March 2006, can be found at the Minor Planet Center website, 
http://cfa-www.harvard.edu/cfa/ps/mpc.html}. 

As one extreme case, in the fast rotator approximation, the time that it
takes to sublimate a layer equal to the penetration-depth of the solar
protons, roughly a $100 \mu m$ dust layer, is only of $\sim 200$yr. 

In the case of the slower rotators we may take, as another extreme case, the
stationary approximation. The stationary Trojan temperature of the side
facing the Sun can be estimated using the equation of effective temperature
with back irradiation from half a sphere, rather than the entire surface of
the sphere as in the case of a fast rotator. This leads to a higher
temperature, namely $T_S \approx 168^o K$, and the corresponding mass-loss
rate $dm/dt(T_S) \approx 3\ 10^{-4} kg\ m^{-2}\ s^{-1}$ (CRC-Handbook of
Chemistry and Physics 1981-1982). In this case, the fraction of the solid
mass that cannot be ejected by gas drag can be estimated as (Jewitt 2002): 
$$f_M=\frac{log(a_M/a_c)}{log(a_M/a_m)}$$ where $a_M$ and $a_m$ are the maximum and
the minimum grain sizes in the distribution respectively and $a_c$, the
maximum particle-radius that is retained, given by:   
$$a_c = \frac{v_e(R)\ {dm/dt(T_S)}}{G\ R\ \rho^2},$$
where $v_e(R)$ is the escape velocity from a body of size $R$ and $G$ is the
gravitational constant. If we take $a_M=1mm$ and $a_m=1 \mu m$ and
$R=100km$, then $a_c \approx 2 \mu m $. We find that, for a body of
rotational period $P=40 hr$, the timescale in which an insulator-mantle of
thickness $L=100 \mu m$ 
grows, is only $\tau_{M}(P=40 hr) \approx 3 days$.

\subsection{Timescale of collisional re-surfacing}

To compute the timescale of the collisional resurfacing process we have
adapted the model by Gil-Hutton (2002) and incorporated the latest
knowledge gained by the Deep Impact mission (see for example
Richardson et al. 2007).

The time to cover the entire surface of an asteroid of radius $R$ with
debris extracted from below the surface by the action of physical
collisions, $\tau_{CR}$, is given by: \[ \tau_{CR} = \frac{1}{\dot{S_r}},\]
where $\dot{S_r} = \dot{S}/{(4 \pi R^2)}$ is the fraction of the surface
that is covered by material excavated by collisions per year and $ \dot{S}$
is the total area covered by collisional ejecta each year.

The total area covered by ejecta per year, $ \dot{S}$, for an object of
radius $R$ is:
\begin{equation}
\label{eq:1}
\dot{S} = \int_{r_{min}}^{r_{max}} \dot{N}(r)\ A_E(r)\ dr,\end{equation}
where $\dot{N}(r)$ is the number of collisions per unit time and
$A_E(r)$ is the area covered by the ejecta produced when projectiles
of radii $r$ collide with an object of radius $R$, at a typical
encounter velocity $v$. We write $A_E$ as $A_E = l \times A_c(R,r,v)$,
where $A_c(R,r,v)$ is the area of the crater produced and $A_c = \pi\
d^2$, where $d$ is the diameter of an idealized crater, and we assume
the densities of the target and the impactor to be equal, i.e. $\rho_R
= \rho_r = 1.5 g cm^{-3} $. The volume of the crater in the gravity dominated cratering
regime is estimated as (Richardson et al. 2007, Holsapple 1993):
\begin{equation}
\label{eq:2}
V_g = K_1\ \left( \frac{m}{\rho} \right) \left( \frac{r\ g(R)}{v^2}
\right)^{\frac{-3 \mu}{2 + \mu}},
\end{equation}
where $g(R)$ is the surface gravity of the target, $m$ is the mass of
the impactor, $v$ its impact velocity, $\rho$ is the density which
is  assumed to be equal for the target and
the impactor. The scaling constants $K_1=0.24$ and $\mu=41$,
correspond to a loosely bound material such as sand, which gave a good
agreement with the size of the crater produced in the Deep impact
mission (Richardson et al. 2007). The diameter of the crater is easily
obtained from $$ V_g = \frac{1}{24} \pi d^3 . $$

Now, we can write $\dot{N}$ as, \[\dot{N} = P_i\ (R+r)^2\ dN(r),\] where
$P_i$ is the intrinsic probability of collisions, as defined by Wetherill
(1967), is the probability of collision per unit time, per unit cross
section and per number of colliding pairs and $dN(r)$ is the number of
objects with radii $r$. The values of $P_i$ and $v$, have been computed
specifically for the Trojans by Dell'Oro \etal(1998) giving approximately
$6.5\ 10^{-18}\ km^{-2}\ yrs^{-1} $ and $5\ km\ s^{-1}$ respectively.

The integral on equation~\ref{eq:1} depends critically on the limits, which
must be handled carefully. The upper limit, $r_{max}$, is the maximum radii
of the projectiles that do not shatter a target of radius $R$. For the
$\rho_R = \rho_r$ case is, \[r_{max} = R\ (5\ \beta v_c^2 - 1)^{-1/3},\]
(Petit \& Farinella 1993), where $\beta=10^{-8}\ g\ erg^{-1}$ is a
crater excavation coefficient. While $r_{min}$ is the cut-off radius.

Finally, to compute $\tau_C$, we need an estimation of the observed
distribution of sizes in the Trojan population, which, for the
smallest objects is (Jewitt et
al. 2000),
\begin{equation}
\label{eq:3}
dN(r) = 1.5\ 10^6\ \left( \frac{1\ km}{r} \right)^3.
\end{equation}

\subsubsection{The choice of $r_{min}$, the cut-off radius}

The contribution of the dust phase, i.e. $\mu m$ sized particles, to the
modification of the a 100 $\mu m$ layer of irradiated material is
negligible, since it would only produce re-arrangements. For such
small impactors, the cratering regime is
strength-dominated.  The volume of the crater produced $V_S$ can be
estimated as (Richardson et al 2007, Holsapple 1993):
$$
V_S = K_1\ \frac{m}{\rho}  \left( \frac{\bar{Y}}{\rho v^2}
\right)^{-\frac{3}{2} \mu},
$$
where $\bar{Y}$ is the effective
yield strengths of the target material. We adopt a value of $\bar{Y}
\approx  10 Pa$, as determined for comet Temple 1 and also in agreement with
laboratory experiments (Richardson et al. 2007,
Bar-Nun et al. 2007). Under the
assumptions used previously, the depth
of this crater for an impactor of size $10 \mu m$ is
$600 \mu m$, i.e., roughly the penetration depth of the solar protons.

Moreover, micro-impacts contribute less to the resurfacing because
dust is carried away by radiation and its size distribution is much
shallower than the one corresponding to macroscopic objects. Therefore
the smallest cut-off radius that we shall assume is $1 mm$, since
objects of this size are not affected by radiation. Unfortunately,
there is no information available on the size distribution of objects
of sizes ranging between $1 mm$ and $100's\ m$ orbiting in the Trojan
clouds. We assume that the size distribution of the smallest observed
Trojans (Jewitt et al. 2000) also applies for small particles.
Although this assumption may seem extreme, there is no creation or
destruction mechanism that affects the small particles differently
than the large objects, since none of them are affected by radiation.
And the evidence of the cratering record on asteroids indicates that
the slope of the size distribution of impactors is unaltered from
$km$-sized bodies down to $m$-sized objects (O'Brien et al. 2006).

\begin{figure}
\epsscale{1.05}
\plottwo{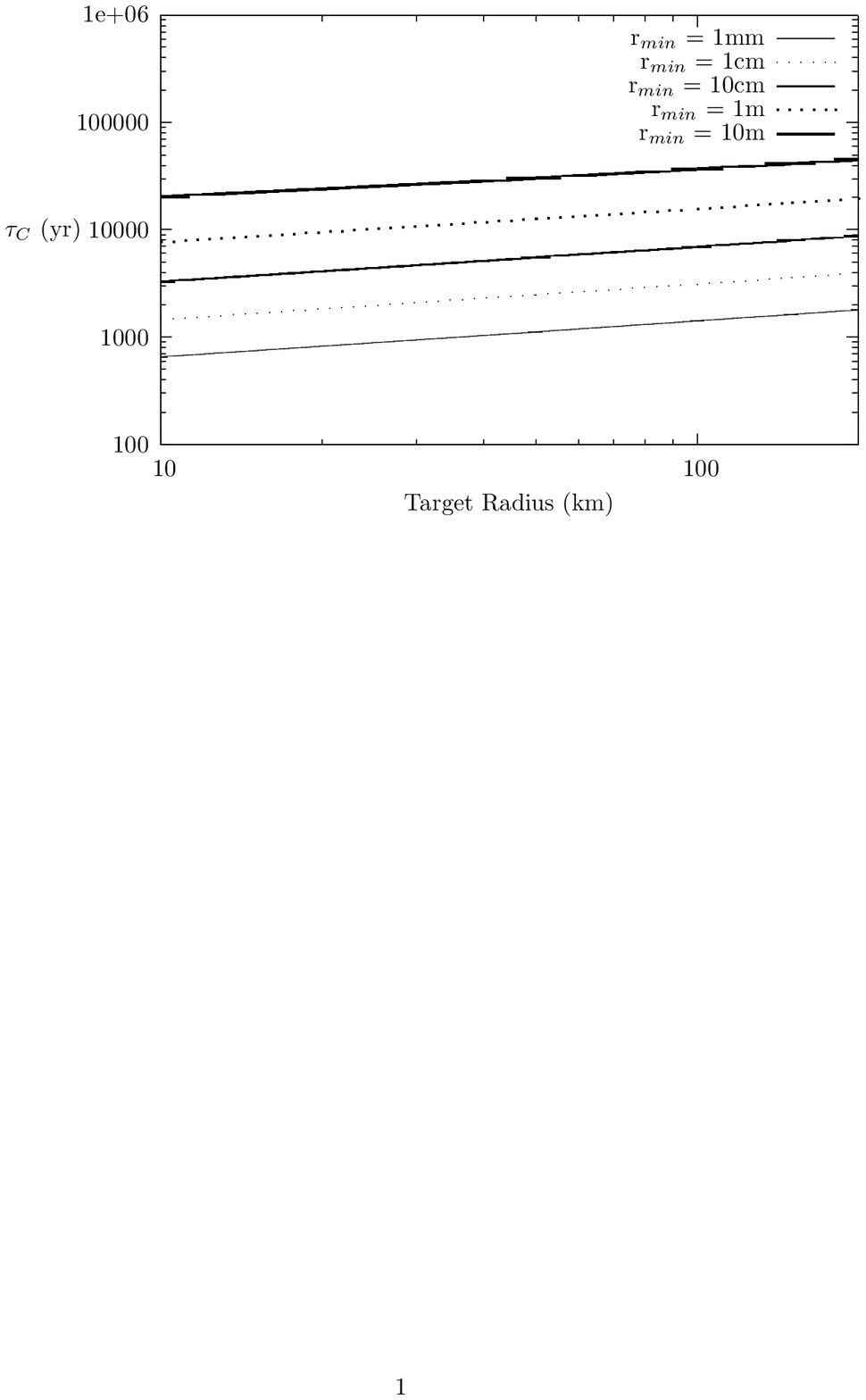}{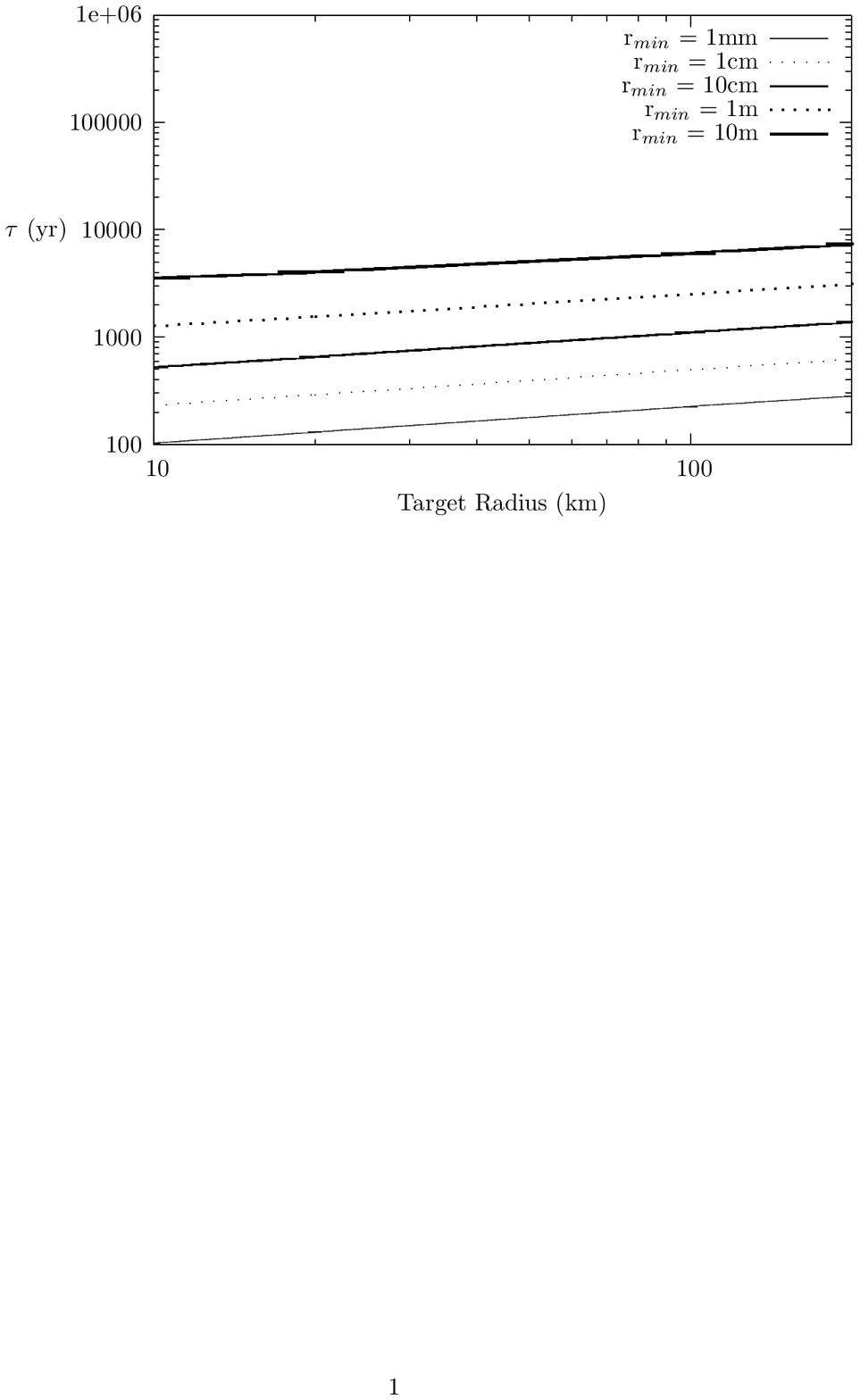}
\caption{Left: Timescale in which the whole body is covered with craters,
$\tau_C$, as a function of the
radius of the target, for the
size range of the known Trojan asteroids. 
Right: Timescale of collisional resurfacing, assuming a value of $l=25$,
$\tau$, as a function of the
radius of the target, for the
size range of the known Trojan asteroids. See text for details.}
\label{fig:3}
\end{figure}

If we take $l=1$, the timescale of collisional resurfacing is,
neglecting overlapping, the time-span in which all the surface is
covered with craters, $\tau_C$. Note that $\tau_C$ is larger than the
time to cover the object with ejecta. Values of $\tau_C$ for the
size range of the known Trojan asteroids, are shown in
figure~\ref{fig:3} for different values of the cut-off radius
$r_{min}$. For a value of $r_{min}=1 mm$ the object is completely
covered by craters in
about $700$ to $1100 yrs$, while for $r_{min}=10 m$, it occurs in
$10^{4}$ to $10^{5} yrs$. 

Now we estimate how the area covered by optically thick
ejecta, scales with the size of the crater. If we assume that the
optically thick
ejecta is close to the edge of the crater, and is mostly originated by
material
close to it, then the ejection velocity from the impact, $v_e$, 
scales as (Richardson et al.
2007): 
$$v_e \propto \sqrt{g\ \frac{d}{2}}.$$ 
Then, for the distance travelled by the ejecta, $h$, is: 
$$h \propto  \frac{v_e^2}{g} = \frac{d}{2},$$
and therefore, the ratio $$l = \left( \frac{h}{d/2} \right)^2$$ 
is constant. To fix
this constant we use the fact that the size of the crater produced in
comet Temple 1 by the Deep Impact mission has a radius of
approximately $30m$ and the base of the plume after the event had a
radius of about $150m$ (Richardson et al.
2007), hence we adopt a value for the ratio of the areas of $l=25$. 
In figure~\ref{fig:3} we also plot values of the timescale for collisional
resurfacing, $\tau$, assuming a value of $l=25$.  For a value of
$r_{min}=1 mm$ the resurfacing occurs in
about $100$ to $300 yrs$, while for $r_{min}=10 m$, it occurs in
$3000$ to $7000 yrs$.

\section{Conclusion}
\label{sec:disc}

The surface properties of a Trojan asteroid are determined by the
interplay of three different mechanisms. When an impact occurs, if it
is sufficiently energetic, the inner materials of the body are exposed
and scattered on the surface. Given that the site of residence and
formation of the Trojans is the outer Solar System, water ice could be
abundant, but, as shown in section
\ref{sec:2.3}, it sublimates fast from the surface, leaving a
mantle of dust, most probably with a red spectroscopic slope. If this
dust layer would remain unaltered for more than $10^{3} yrs$, its
spectroscopic slope would turn to neutral by the action of solar
protons. But impacts occur so frequently that the irradiation mantle
is disrupted. 

The surfaces of most Trojans are red, but some neutral objects exists,
therefore, it is suggested that the irradiation resurfacing timescale is
similar, but not much shorter, than the collisional one. The conclusion
reached here is that the timescale to fill the entire surface of the Trojan
asteroids with craters is similar to that of the flattening of the red
spectra by solar-proton irradiation, if the cut-off radius of the size
distribution is $1 mm$ and its slope is $-3$ -for the whole size-range. If
the contribution of ejecta is taken into account, a cut-off size of $1 m$,
with the same slope, renders a collisional resurfacing-timescale in the
required range. 

The model presented here can be seen as an exercise to illustrate a scenario
of collisional and radiation balance to understand the distribution of
spectral slopes in the Trojan swarms. Naturally, refinements are needed to
represent better reality. For example, our estimates of the collisional
resurfacing timescales bear large uncertainties, mainly due to our lack of
knowledge of the size distribution of small particles around the Lagrangian
points of the orbit of Jupiter. On the other hand, the information regarding
visual spectral slopes of known dynamical-family members (Fornasier et al.
2007), indicates that most dynamical families are remarkable uniform. With
the noticeable exception of the one of Eurybathes, they are mostly
moderately red, suggesting a correlation with age. Eventually, the
calculations presented here could be used to understand such a potential
trend.  

Finally, a comment on how the scenario would change if the mass-loss rate of
water ice at $120^oK$, based in realistic rotation rates and albedos where
to be higher than estimated here. In this case mass loss due to sublimation
would be higher and the exposed surface ice could have been lost from the
Trojans in their current orbit. A different orbit is not required.   

\acknowledgements

\section*{Acknowledgements}

We are grateful for the comments of an anonymous referee, which have helped
to improve this article greatly, particularly regarding the sublimation rate
of water ice. 

GS has been supported by Italian Space Agency contract n. I/015/07/0
(Studi di Esplorazione Sistema Solare). A.B-N acknowledges partial
support by the US-Israel Binational Science foundation. MDM
acknowledges partial support by UBACyT X465. \\

\section*{References}

Bar-Nun, A., Dror, J., Kochavi, E., Laufer, D. 1987. Amorphous water ice
and its ability to trap gases. {\it Physical Review B}, {\bf 35}, 2427-2435.

Bar-Nun, A., Pat-El, I., Laufer, D. 2007. Comparison between the
findings of Deep Impact and our experimental results on large samples
of gas-laden amorphous ice. {\it Icarus}, {\bf 191}, 2, 562-566.

Beauge, C., Roig, F. 2001. A Semi-analytical Model for the Motion of the
Trojan Asteroids: Proper Elements and Families. {\it Icarus}, {\bf 153}, 2,
391-415.

%Bus, S.J. 1999. Compositional structure in the asteroid belt: results of a
%spectroscopic survey. {\it PhD Thesis}, MIT, Cambridge.

Bendoya.,P, Cellino, A, Di Martino, M., Saba, L. 2004. Spectroscopic
observations of Jupiter Trojans. {\it Icarus}, {\bf 168}, 2, 374-384.

Binzel, R.P. and  Sauter, L.M. 1992.
Trojan, Hilda, and Cybele asteroids - New lightcurve observations and
analysis.
{\it Icarus}, {\bf 95}, 222-238.

Brunetto, R., Roush, T.L. 2008. Impact of irradiated methane ice crusts on
compositional interpretations of TNOs. {\it Astronomy and Astrophysics},
{\bf 481}, 879-882.

%Chambers, J.E. 1999. A Hybrid Symplectic Integrator that Permits Close
%Encounters between Massive Bodies. {\it M.N.R.A.S.}, {\bf 304}, 793-799.

CRC-Handbook of Chemistry and Physics, 1981-1982, The Chemical Rubber 
Co.Cleveland OH.

Cruikshank, D. P., Dalle Ore, C. M., Roush, T. L., Geballe,
T. R., Owen, T. C., de Bergh, C., Cash, M. D.;
Hartmann, W. K. 2001. Constraints on the Composition of Trojan
Asteroid 624 Hektor. {\it Icarus}, {\bf 153}, Issue 2, 348-360.

Dell'Oro, A., Marzari, P., Paolicchi F., Dotto, E., Vanzani, V. 1998. Trojan
collision probability: a statistical approach. {\it Astronomy and
Astrophysics}, {\bf 339}, 272-277.

%Dotto, E., Fornasier, S., Barucci, M. A., Licandro, J., Boehnhardt, H.,
%Hainaut, O., Marzari, F., De Bergh, C., De Luise, F. 2006. The surface
%composition of Jupiter Trojans: Visible and Infrared survey of dynamical
%Families. {\it Icarus}, In Press.

%Dvorack, R., Tsiganis, K. 2000. Why do Trojan ASCS (not) escape?.
%{\it Celestial
%Mechanics and Dynamical Astronomy}, {\bf 78}, 125-136.

Emery, J.~P. , Brown, R.~H. 2003.
Constraints on the surface composition of Trojan asteroids
from near-infrared (0.8-4.0 {$\mu$}m) spectroscopy,
{\it Icarus}, {\bf 164}, {104-121}.

Emery, J.~P. , Brown, R.~H. 2004.
The surface composition of Trojan asteroids: constraints set by
scattering theory. {\it Icarus}, {\bf 170}, {131-152}.

Emery J.P., Cruikshank D.P. , Van Cleve J. 2006. Thermal emission
spectroscopy (5.2-38 $\mu$m) of three Trojan asteroids with the
Spitzer Space Telescope: Detection of fine-grained silicates. {\it
Icarus}, {\bf 82 }, 2, 496-512.

Fern{\' a}ndez, Y.~R. , Sheppard, S.~S. , Jewitt, D.~C. 2003.  The
Albedo Distribution of Jovian Trojan Asteroids. {\it The Astronomical Journal},
{\bf 126}, 1563-1574.

Fornasier, S., Dotto, E., Hainaut, O., Marzari, F., Boehnhardt, H., de
Luise, F., Barucci, M. A. 2007. Visible spectroscopic and photometric survey
of Jupiter Trojans: Final results on dynamical families. {\it Icarus} {\bf
190}, 2, 622-642. 

%Fornasier, S. and Dotto, E. and Marzari, F. and Barucci, M.~A. and
%Boehnhardt, H. and Hainaut, O. and de Bergh, C., 2004, Visible spectroscopic
%and photometric survey of L5 Trojans: investigation of dynamical families,
%{\it Icarus}, {\bf 172}, 221-232.

Gil-Hutton, R. 2002. Color diversity among Kuiper belt objects: The
collisional resurfacing model revisited. {\it Planetary and Space Science}, {\bf
50}, 1, 57-62.

Gosling, J.T., 2007, The Solar wind, in: Encyclopedia of the Solar System
2nd Edition, L.McFadden, P.R. Weissman, T.V. Johnson Eds, Academic Press, pp
99-116

%Horner J, Evans N.W. and Bailey M.E. 2005. Simulation of the Population of
%Centaurs II. Individual objects. {\it
%M.N.R.A.S.}, {\bf 355}, 2, 321-329.

Hartmann, W. K., Binzel, R.P., Tholen, D.J., Cruikshank, D. P., Goguen, J.
1988. Trojan and Hilda asteroid light-curves. I - Anomalously elongated
shapes among Trojans (and Hildas?). {\it Icarus}, {\bf 73}, 487-498.

Holsapple K.A. 1993. The scaling of impact processes in planetary
sciences. {\it Annu. Rev. Earth Planet. Sci.}, {\bf 21}, 333-374.

Hudson R.L., Palumbo M.E., Strazzulla G., Moore M.H., Cooper J.F., , Sturner
S.J., 2008, Laboratory studies of the chemistry of TNO surface materials,
in: The Solar System beyond Neptune, The University of Arizona Space Science
Series, M. A. Barucci, H. Boehnhardt, D. P. Cruikshank, A. Morbidelli
(editors), University of Arizona Press, Tucson, pp 507-523

Jewitt, D.C. 2002, From Kuiper Belt Object to Cometary Nucleus: The Missing
Ultrared Matter. {\it The Astronomical Journal}, {\bf 123}, 2,
1039-1049.

Jewitt, David C., Luu, Jane X. 1990. CCD spectra of asteroids. II - The
Trojans as spectral analogs of cometary nuclei. {\it Astronomical Journal},
{\bf 100}, 933-944.

Jewitt, David C., Trujillo, Chadwick A., Luu, Jane X. 2000.
Population and Size Distribution of Small Jovian Trojan Asteroids. 
{\it The Astronomical Journal}, {\bf 120}, 2, 1140-1147.

%Karlsson, O. 2004. Transitional and temporary objects in the Jupiter Trojan
%area. {\it Astronomy and Astrophysics}, {\bf 413}, 1153-1161.

%Landolt, A. U. 1992.
%UBVRI photometric standard stars in the magnitude range 11.5-16.0 around the
%celestial equator.
%{\it Astr. J.} {\bf 104}, 1, 340-371, 436-491.

Lanzerotti L. J., Brown, W. L., Poate, C. M.,  Augustyniak, W. M.
1978. Low energy cosmic ray erosion of ice grains in interplanetary
and interstellar media. {\it Nature} {\bf 272}, 431-433.

%Lazzaro, D., et al.: 2004. S3OS2: The visible spectroscopic survey of 820
%asteroids. {\it Icarus},   {\bf 172}, 179--220.

%Levison, H., Shoemaker, E. M., Shoemaker, C. S. 1997. The dispersal of the
%Trojan asteroid swarm. {\bf Nature}, {\bf  385},  42-44.

%Marzari, F., Farinella, P., Vanzani, V.
%Are Trojan collisional families a source for short-period comets?. 1995.
%{\it Astronomy and Astrophysics}, {\bf 299}, 267.

Marzari, F., Farinella, P., Davis, D. R., Scholl, H., Campo Bagatin, A.
 1997.  {\it Icarus}, {\bf 125}, 1,  39-49.

%Marzari, F. and Scholl, H. 1998. Capture of Trojans by a Growing
%Proto-Jupiter {\it Icarus}, {\bf 131}, 1, 41-51.

Lisse, C. M., VanCleve, J., Adams, A. C., A'Hearn, M. F., Fernández, Y. R.,
Farnham, T. L., Armus, L., Grillmair, C. J., Ingalls, J., Belton, M. J. S.,
Groussin, O., McFadden, L. A., Meech, K. J., Schultz, P. H., Clark, B. C.,
Feaga, L. M., Sunshine, J. M. 2006.
Spitzer Spectral Observations of the Deep Impact Ejecta.
{\it Science}, {\bf 313}, 5787, 635-640.

Mayer, E., and Pelzer, R., 1986. Astrophysical implications of amorphous ice
- A microporous solid. {\it Nature}, {\bf 319}, 298-301.

Melita, M.D., Licandro, J., Jones, D., Wiliams, I.P. 2008a. Physical
properties and orbital stability of the Trojan asteroids.
{\it Icarus}, {\bf 195}, 2, 686-697.

Melita, M.D., Duffard, R., Williams I.P., Jones, D.C., Licandro, J., Ortiz
J.L. 2008b. Lightcurves of 21 Trojan Asteroids. {\it A \& A}, Submitted.

%Milani, A., Nobili, A.M. 1992. An example of stable chaos in the Solar
%System {\it Nature}. {\bf 357}, 6379, 569-571.

%Milani, A. 1993.  The Trojan asteroid belt: Proper elements, stability,
%chaos and families {\it Celestial Mechanics and Dynamical Astronomy}, {\bf
%57}, 1-2, 59-94.

%Milani, A., Nobili, A.M., Knezevic, Z. 1997.  Stable Chaos in the Asteroid Belt.
%{\it Icarus}, {\bf 125}, 1, 13-31.

%Morbidelli, A., Levison, H. F., Tsiganis, K. and Gomes, R. 2005.
%Chaotic capture of Jupiter's Trojan asteroids in the early Solar System
%{\it Nature}, {\bf 435}, 7041, 462-465.

Moroz, L. V., Baratta, G., Strazzulla,G. , Starukhina,L. V., Dotto,
E., Barucci, . M.A., Arnold, G.,  Distefano, E., 2004. Optical
alteration of complex organics induced by ion irradiation: I.
Laboratory experiments suggest unusual space weathering trend. {\it
Icarus} {\bf 170}, 214-228

%Nesvorny, D. and Dones, L. 2002. How Long-Lived Are the Hypothetical Trojan
%Populations of Saturn, Uranus, and Neptune? {\it Icarus}, {\bf 160}, 2,
%271-288.

Notesco, G., Bar-Nun, A., Owen, T. 2003. Gas trapping in water ice at
very low deposition rates and implications for comets. {\it Icarus},
{\bf 162}, 1, 183-189.

O'Brien, D.P., Greenberg, R., Richardson, J.E. 2006.  Craters on
asteroids: Reconciling diverse impact records with a common impacting
population. {\it Icarus}, {\bf 183}, 1, 79-92.

Petit, J.M. , Farinella, P. 1993.  Modelling the outcomes of
high-velocity impacts between small solar system bodies. {\it Celestial
Mechanics and Dynamical Astronomy}, {\bf 57}, 1-2, 1-28.

%Szabo, Gy. M., Ivezic, E., Juric, M., Lupton, R.
%The properties of Jovian Trojan asteroids listed in SDSS Moving Object
%Catalogue 3. 2007. {\it MNRAS}, {\bf 377}, 4, 1393-1406. 

%Pilat-Lohinger, E.,  Dvorak, R. 1999. {\it Celestial
%Mechanics and Dynamical Astronomy}, {\bf 73}, 117-126.

%Rabe, E. 1972. Orbital Characteristics of Comets Passing Through the 1:1
%Commensurability with Jupiter. The Motion, Evolution of Orbits, and Origin
%of Comets, Proceedings from IAU Symposium no. 45, held in Leningrad,
%U.S.S.R., August 4-11, 1970. Edited by Gleb Aleksandrovich Chebotarev, E. I.
%Kazimirchak-Polonskaia, and B. G. Marsden. International Astronomical Union.
%Symposium no. 45, Dordrecht, Reidel, p.55.

%Strazzulla G. Johnson R.E. 1991. Iradiation effects on comets and cometary
%debris. In: Newburn Jr. R.L. Neugebauer, M. Rahe J. (Eds), Comets in the
%post-Halley era. Kluwer Academic publishers. 243-275.

%Szevehely, V. 1967. Theory of orbits: The restricted Problem of three
%bodies. Academic Press Inc. New York.

%Tedesco, E.F., Noah, P.V., Noah, M., Price and Stephan D. 2002.
%The Supplemental IRAS Minor Planet Survey.
%{\it The Astronomical Journal}, {\bf 123}, 2, 1056-1085.

%Tsiganis, K., Dvorak, R., Pilat-Lohinger, E. 2000.
%Thersites: a jumping' Trojan?
%{\it Astronomy and Astrophysics}, {\bf 354}, 1091-1100.

%Tsiganis, K., Varvoglis, H., Dvorak, R. 2005.  Chaotic Diffusion And
%Effective Stability of Jupiter Trojans {\it Celestial Mechanics and
%Dynamical Astronomy}, {\bf 92}, 1-3. 71-87

%Yoder, C. F. 1979. Notes on the origin of the Trojan asteroids. {\it Icarus},
%{\bf  40}, 341-344.

Richardson, J.E., Melosh, H.J., Lisse, C.M., Carcich, B. 2007. A
ballistics analysis of the Deep Impact ejecta plume: Determining Comet
Tempel 1's gravity, mass, and density.  {\it Icarus}, {\bf 190}, 2,
357-390.

Sack, N.J. and Baragiola, R.A., 1993. Sublimation of vapor-deposited water
ice below 170 K, and its dependence on growth conditions. {\it Phys. Rev. B},
{\bf 48}, 9973 – 9978.

Yang, B, Jewitt, D. 2007. Spectroscopic Search for Water Ice on Jovian
Trojan Asteroids. {\it The Astronomical Journal}, {\bf 134}, 1,
223-228.

Wetherill, G. W. 1967.  Collisions in the asteroid belt. 
{\it J. Geophys. Res.}, {\bf 72}, 2429-2444.

Ziegler, J. F., Biersack, J. P., and Littmark, U., 1996. The Stopping
and Range of Ions in Solids, Pergamon Press, New York.

\newpage

%\begin{figure}[bp]
%\centerline{\includegraphics[width=16cm,height=12cm]{histoS2.ps}}
%\caption{Histogram of the slopes of visual reflection spectra
%normalized at $6000 \AA$ for the Trojan asteroids. The sources of this
%data can be found in Melita et al. 2008.}
%\label{fig:1}
%\end{figure}

\newpage

%\begin{figure}[bp]
%\centerline{\includegraphics[width=16cm,height=12cm]{SvsD.ps}}
%\caption{Slopes of visual reflection spectra normalized
%at $6000 \AA$ vs. D(0.04), the physical diameter, assuming an albedo
%of 0.04, for the Trojan asteroids.
%Data taken form Melita et al. 2008.}
%\label{fig:2}
%\end{figure}

\newpage

%\begin{figure}[bp]
%\centerline{\includegraphics[width=16cm,height=12cm]{coll6.ps}}
%\caption{Timescale in which the whole body is covered with craters,
%$\tau_C$, as a function of the
%radius of the target,
%for the
%size range of the known Trojan asteroids,}
%\label{fig:3}
%\end{figure}

\end{document}